\documentclass[sigconf]{acmart}

\AtBeginDocument{%
  }

\copyrightyear{2026}
\acmYear{2026}
\setcopyright{cc}
\setcctype{by}
\acmConference[LAK 2026]{LAK26: 16th International Learning Analytics and Knowledge Conference}{April 27-May 01, 2026}{Bergen, Norway}
\acmBooktitle{LAK26: 16th International Learning Analytics and Knowledge Conference (LAK 2026), April 27-May 01, 2026, Bergen, Norway}
\acmPrice{}
\acmDOI{10.1145/3785022.3785092}
\acmISBN{979-8-4007-2066-6/2026/04}

\usepackage{graphicx}
\usepackage{booktabs}
\usepackage{xcolor}
\usepackage{listings}
\usepackage{multirow}
\usepackage{threeparttable}
\usepackage{float}
\usepackage{pgfplots}
\pgfplotsset{compat=1.18}
\usepackage{siunitx}

\newcommand{\osfRepo}{\href{https://osf.io/f2dvk/overview?view_only=16e581c26c8b4d1c9b3a1c6b06b89399}{osf.io/f2dvk}}

\title[\emph{MedSimAI}]{\emph{MedSimAI}: Simulation and Formative Feedback Generation to Enhance Deliberate Practice in Medical Education}

\author{Yann Hicke}
\affiliation{%
  \institution{Cornell University}
  \city{Ithaca}
  \state{NY}
  \country{USA}}

\author{Jadon Geathers}
\affiliation{%
  \institution{Cornell University}
  \city{Ithaca}
  \state{NY}
  \country{USA}}

\author{Kellen Vu}
\affiliation{%
  \institution{Weill Cornell Medicine}
  \city{New York}
  \state{NY}
  \country{USA}}

\author{Justin Sewell}
\affiliation{%
  \institution{UCSF School of Medicine}
  \city{San Francisco}
  \state{CA}
  \country{USA}}

\author{Claire Cardie}
\affiliation{%
  \institution{Cornell University}
  \city{Ithaca}
  \state{NY}
  \country{USA}}

\author{Jaideep Talwalkar}
\affiliation{%
  \institution{Yale School of Medicine}
  \city{New Haven}
  \state{CT}
  \country{USA}}

\author{Dennis Shung}
\affiliation{%
  \institution{Mayo Clinic}
  \city{Rochester}
  \state{MN}
  \country{USA}}

\author{Anyanate Gwendolyne Jack}
\affiliation{%
  \institution{Weill Cornell Medicine}
  \city{New York}
  \state{NY}
  \country{USA}}

\author{Susannah Cornes}
\affiliation{%
  \institution{UCSF School of Medicine}
  \city{San Francisco}
  \state{CA}
  \country{USA}}

\author{Mackenzi Preston}
\affiliation{%
  \institution{Weill Cornell Medicine}
  \city{New York}
  \state{NY}
  \country{USA}}

\author{Rene Kizilcec}
\affiliation{%
  \institution{Cornell University}
  \city{Ithaca}
  \state{NY}
  \country{USA}}

\renewcommand{\shorttitle}{\emph{MedSimAI}: Simulation \& Feedback for Deliberate Practice}
\renewcommand{\shortauthors}{Hicke et al.}

\begin{document}
\begin{abstract}
Medical education faces challenges in providing scalable, consistent clinical skills training. Simulation with standardized patients (SPs) develops communication and diagnostic skills, but remains resource-intensive and variable in feedback quality. Existing AI-based tools show promise yet often lack comprehensive assessment frameworks, evidence of clinical impact, and integration of self-regulated learning (SRL) principles.  
Through a multi-phase co-design process with medical education experts, we developed \emph{MedSimAI}, an AI-powered simulation platform that enables deliberate practice through interactive patient encounters with immediate, structured feedback. Leveraging large language models, \emph{MedSimAI} generates realistic clinical interactions and provides automated assessments aligned with validated evaluation frameworks.  
In a multi-institutional deployment (410 students; 1{,}024 encounters across three medical schools), 59.5\% engaged in repeated practice. At one site, mean Objective Structured Clinical Examination (OSCE) history-taking scores rose from 82.8 to 88.8 ($p<0.001$, $d=0.75$), while a second site’s pilot showed no significant change. Automated scoring achieved 87\% accuracy in identifying proficiency thresholds on the Master Interview Rating Scale (MIRS). Mixed-effects analyses revealed institution and case effects. Thematic analysis of 840 learner reflections highlighted challenges in missed items, organization, review-of-systems, and empathy.  
These findings position \emph{MedSimAI} as a scalable formative platform for history-taking and communication, motivating staged curriculum integration and realism enhancements for advanced learners.
\end{abstract}



\maketitle

\section{Introduction}
Large Language Models (LLMs) are transforming educational practices across disciplines—from mathematics instruction \cite{pardos2024chatgpt} and interactive, dialogue-based tutoring \cite{nye2023generative} to programming education \cite{phung2023generating}. In medical education, where rigorous, long-term training and repeated clinical practice are essential, AI-powered simulations offer significant opportunities to enhance physician preparation. By leveraging LLMs' ability to generate realistic, interactive learning experiences, educators can create scalable training platforms that provide immediate, personalized feedback while accommodating diverse clinical scenarios.

Medical education represents one of the most rigorous training pathways in professional education. For example, in the United States, students complete four years of undergraduate medical education, mastering foundational sciences and clinical skills, before progressing to 3–7 years of residency and potential fellowship training (models of medical education certainly vary internationally). Throughout this journey, future physicians must acquire both vast knowledge and clinical skills—from patient interviewing to complex decision-making—while maintaining quality care. Accordingly, training relies on competency-based assessments: learners progress from practicing clinical skills with feedback in simulation to caring for real patients under supervision \cite{govaerts2013validity,cutrer2017fostering}.

Educators implement this feedback-driven approach through simulation-based learning, which has gained widespread traction. It enables educators to assess and reinforce both technical and interpersonal skills by recreating clinical scenarios with realistic conditions, including with mannequin simulators, standardized patients (trained actors who portray patients with specific conditions), or virtual simulations \cite{talwalkar2020twelve,uchida2023standardized}. The effectiveness of this approach is well-documented \cite{beal2017effectiveness}, and over 80\% of U.S. medical schools report using simulation-based learning \cite{passiment2011medical,hayden2010use,vyas2013use}. Standardized patients play a particularly central role in assessment, appearing in 79\% of formative and 91\% of summative exams prior to graduation \cite{uchida2019approaches}.

Despite its central role, traditional simulation-based training faces critical challenges. High-fidelity mannequins cost tens of thousands of dollars, creating accessibility barriers \cite{elendu2024impact}. Current methods lack diversity in patient representation, leaving gaps in culturally competent care training \cite{foronda2020underrepresentation,ross2021key}. Furthermore, while effective feedback is crucial for learning, delivering high-quality, timely debriefing remains challenging due to limited instructor resources \cite{ismail2024challenges,harrison2015barriers}.

Recent learning-science research offers insights into how AI-powered simulations can address these limitations. First, AI-based adaptive feedback can outperform static expert solutions in improving diagnostic justification quality \cite{Bauer2025Adaptive}. Second, personalizing simulation parameters—case difficulty, feedback timing, and prompt wording—to individual learner characteristics offers a scalable path toward learner-centered training that traditional methods cannot achieve \cite{Bauer2025Personalizing}. Finally, learners' perceived authenticity and cognitive load during simulation reliably predict downstream diagnostic accuracy across domains \cite{Chernikova2024Relation}, suggesting that well-designed AI simulations could enhance learning outcomes.

Current AI implementations in medical education show promise but only partially realize this potential. AI-simulated patients can improve clinical interviewing skills \cite{yamamoto2024enhancing}, and LLM-powered assessment shows consistency with human raters \cite{holderried2024language}. However, existing AI-based medical training tools have critical limitations. Most address only narrow competencies, lack comprehensive assessment frameworks, offer limited evidence of downstream clinical impact, and rarely incorporate self-regulated learning (SRL) scaffolds—despite extensive evidence that SRL abilities predict stronger diagnostic reasoning, improved exam performance, and sustained professional development in medical trainees \cite{SandarsCleary2011,BrydgesButler2012,Li2023}.

We developed \emph{MedSimAI} to address these limitations through a comprehensive AI-powered simulation platform designed to meet diverse educational needs while improving accessibility. Building on the learning-science insights above, \emph{MedSimAI} embeds adaptive, SRL-aligned feedback loops and instructor-configurable case fidelity—features that address the scalability and personalization challenges inherent in traditional simulation-based training. Beyond reporting outcomes, we contribute multi-site platform telemetry and an inductive thematic analysis of 840 learner reflections (from 1{,}022 captured), using mixed-effects models to probe case/institution effects and usage–performance relations that inform design and implementation.

Our contributions include:
\begin{enumerate}
    \item Multi-phase co-design with medical education experts at three institutions, yielding an AI-standardized patient platform with instructor-authored cases and configurable realism.
    \item A flexible assessment layer combining multi-rubric scoring (including MIRS) and case-specific checklists to generate quote-grounded formative feedback and SRL-oriented dashboards.
    \item A multi-site evaluation (1{,}024 encounters; 410 learners) combining platform telemetry, mixed-effects analyses, surveys, and an inductive thematic analysis of 840 reflections.
    \item Evidence of downstream impact and validity: a quasi-experimental Objective Structured Clinical Examination (OSCE) history-taking improvement at one institution and independent benchmarking of automated MIRS scoring against trained human evaluators.
\end{enumerate}

\paragraph{Research questions.}
We address four research questions:
\begin{enumerate}
    \item \textbf{RQ1:} How does engagement vary across institutions, cases, and modalities?
    \item \textbf{RQ2:} How do platform-scored outcomes vary across institutions and cases, and how are they associated with within-session behavior and practice dose?
    \item \textbf{RQ3:} How closely does automated scoring align with independent human MIRS ratings?
    \item \textbf{RQ4:} What needs and challenges do learners surface in surveys and reflections, and how do these inform redesign and implementation?
\end{enumerate}

Recent reports further underscore both the promise and the boundary conditions of AI-standardized patients. A curriculum-embedded deployment of an AI patient-actor app highlighted high learner acceptability, increased comfort with generative AI, and the value of low-pressure, feedback-rich practice \cite{thesen2025generative}. Complementing this, a single-blind, two-site randomized controlled trial of AI-standardized patient exam-preparation sessions showed a modest but statistically significant OSCE improvement for the intervention group \cite{lavigne2025ai}. These results motivate our focus on scalable simulation with immediate formative feedback while probing how implementation choices (e.g., integration vs.\ optional use) shape downstream performance.

\section{Related work}
\sloppy
\subsection{Simulated Humans with LLMs}
LLMs have demonstrated a remarkable ability to approximate human behaviour in interactive scenarios—from simulating daily routines \cite{park2023generative} to exhibiting stable personality traits \cite{hilliard2024eliciting}. Although these advances suggest LLMs’ potential as believable proxies for humans, most current approaches remain domain-agnostic. Building socially intelligent AI therefore requires context-specific adaptation \cite{mathur-etal-2024-advancing}. Anthropomorphic cues are essential for believable simulation, yet we must avoid conflating surface-level human-like language with genuine understanding; Shanahan warns that LLMs can appear more sentient than they are \cite{shanahan2023role}. Medical training, in particular, demands specialised AI behaviour—accurate clinical knowledge and structured feedback. We address these needs by extending LLM-based role-play into healthcare simulation.

\subsection{Simulation-based training with LLMs}
Early work on pedagogical agents showed that lifelike virtual characters
can enhance learner engagement and performance \cite{LesterEtAl1997},
and recent LLM systems have built on this foundation to create richer training
environments. In particular, multiple studies leverage LLMs to create virtual training environments. Shorey et al.\ introduced avatar-based patients to help nursing students practice communication skills efficiently \cite{shorey2019virtual}, and Yamamoto et al.\ reported improved exam scores in students who interviewed AI-simulated patients \cite{yamamoto2024enhancing}. Beyond basic interviewing, Wang et al.\ proposed PATIENT-$\Psi$ for mental-health scenarios grounded in cognitive-behavioral therapy \cite{wang2024patient}, and Louie et al.\ showed how expert-crafted principles can produce more realistic AI patients \cite{louie-etal-2024-roleplay}. Others focus on immersive VR  \cite{seo2023development} or just-in-time feedback tools \cite{lin2024imbue}. Separately, a single-blind randomized controlled trial by Desplanque et al.\ found that two months of AI-standardized patient exam-preparation sessions produced statistically significant, albeit modest, OSCE improvements and reduced pre-exam stress \cite{desplanque2025ai}. Our work draws inspiration from this literature and contributes to it: we co-designed \emph{MedSimAI} with faculty experts across three medical schools to support a wider instructional spectrum (history-taking, diagnostic reasoning, and shared decision-making) with instructor-authored cases, multi-rubric analytics, and self-regulated-learning scaffolds. By developing across multiple sites from the start, we set out to build a tool that will generalize across sites.

\subsection{Educational conversational analysis}
Effective AI-driven simulation-based instruction also requires robust dialogue analytics. Jain et al.\ advocated perspective-based summaries to guide deeper reflection \cite{jain-etal-2023-summarize}. In K–12 contexts, Wang et al.\ found that Tutor CoPilot boosts learning outcomes by modeling expert thinking in real time \cite{wang2024tutor}, and their Edu-ConvoKit facilitates systematic processing of educational dialogues \cite{wang2024convokit}. Demszky et al.\ demonstrated that automated feedback can improve online teaching quality \cite{demszky2023m}, and Borchers et al.\ emphasized facilitative peer-tutor dialogue for better learning \cite{borchers2024combining}. Complementary work on NLP-enabled formative feedback shows that structured AI comments drive substantive revisions in students’ science writing \cite{KimEtAl2024} and that feedback accuracy mediates learning gains \cite{KarizakiEtAl2024}. While many dialogue-analytics systems are evaluated in general or K–12 contexts, less is known about their validity and implementation constraints in high-stakes, domain-specific settings such as clinical interviewing. Our work adapts these analytic methods to AI-simulated clinical encounters, offering targeted feedback on both communication and clinical skills.

\subsection{Formative feedback and feedback uptake}
Formative feedback is most effective when it is timely, specific, and oriented toward actionable next steps rather than judgement alone \cite{shute2008focus,nicol2006formative}. However, learners often underuse available feedback unless they have opportunities and support to interpret and act on it; this perspective is captured in work on feedback literacy and uptake \cite{carless2018feedback}. In simulation-based learning, high-quality debriefing operationalizes these principles but can be difficult to scale consistently. \emph{MedSimAI} therefore positions automated scores as formative (not summative), pairs item-level, quote-grounded feedback with SRL scaffolds (goal setting and reflection), and uses clinically authentic metaphors to reduce friction and support follow-through.

\subsection{Self-Regulated Learning in Learning Technologies}
Self-regulated learning (SRL) highlights planning, monitoring, and reflection as crucial to skill development. Kumar et al.\ showed that LLM-based reflective prompts boost student confidence and exam performance \cite{kumar2024supporting}, echoing findings that metacognitive prompts improve learning in computer-based environments \cite{guo2022using}. Yu et al.\ and Cho et al.\ underscored SRL’s importance in online education and clinical contexts \cite{yu2023self}, while Zhang et al.\ employed LLM embeddings to detect SRL strategies in think-aloud data \cite{zhang2024using}. Medical-education research further stresses that deliberate practice supported by SRL processes underpins communication-skill mastery \cite{BrydgesButler2012}. Despite these advantages, few platforms embed SRL scaffolds into real-time clinical simulations. We integrate SRL principles within our LLM-driven environment, ensuring that learners can iteratively practice core competencies, receive immediate feedback, and self-reflect on performance. Beyond proof-of-concepts, early deployments and trials are beginning to quantify impact. In a curriculum-embedded setting, an AI patient-actor app yielded high acceptability and perceived learning value, emphasizing the role of safe, feedback-rich practice opportunities \cite{thesen2025generative}. In parallel, a single-blind, two-site randomized controlled trial found that two months of AI-standardized patient exam-preparation sessions produced a small but significant OSCE gain \cite{lavigne2025ai}. Together, these studies situate AI-mediated practice as a credible adjunct to traditional preparation while leaving open questions about effect size, durability, and the importance of curricular integration—questions our multi-site deployment also probes.

\subsection{Learning analytics in health professions}
Within medical, nursing, and allied-health education, a recent scoping review synthesizes how learning analytics (LA) is currently used, outlining an LA life-cycle and highlighting challenges around validity, feedback actionability, and curricular integration \citep{bojic2023laHPE}. In parallel, “practice-analytics” research examines how clinicians make sense of performance data and how design factors—communication, data quality, infrastructure, and affective risks—shape reflection and behavior change \citep{whitelockwainwright2022sensemaking}. Mixed-method and case-study work further explores clinician sensemaking with dashboards and repurposed clinical data for continuing professional development \citep{whitelockwainwright2024mmcs,whitelockwainwright2024repurposed}. These strands motivate our Learning Hub design: surface rubric-aligned excerpts and case-level trends (rather than global scores alone), communicate uncertainty, and frame planning artifacts in familiar clinical metaphors to reduce cognitive load and support action.

\paragraph{Summary.}
Together, prior work motivates an LA-informed simulation loop in which rich interaction data are transformed into interpretable measures and actionable feedback, then translated into learner and instructor actions. Our contribution is to operationalize this loop end-to-end for clinical interviewing: co-designing a configurable AI-standardized patient, pairing it with rubric/checklist-based feedback, and evaluating deployment and validity across multiple institutions.

\section{Co-Design Process and Insights}

\begin{figure}
    \centering
    \includegraphics[width=\linewidth]{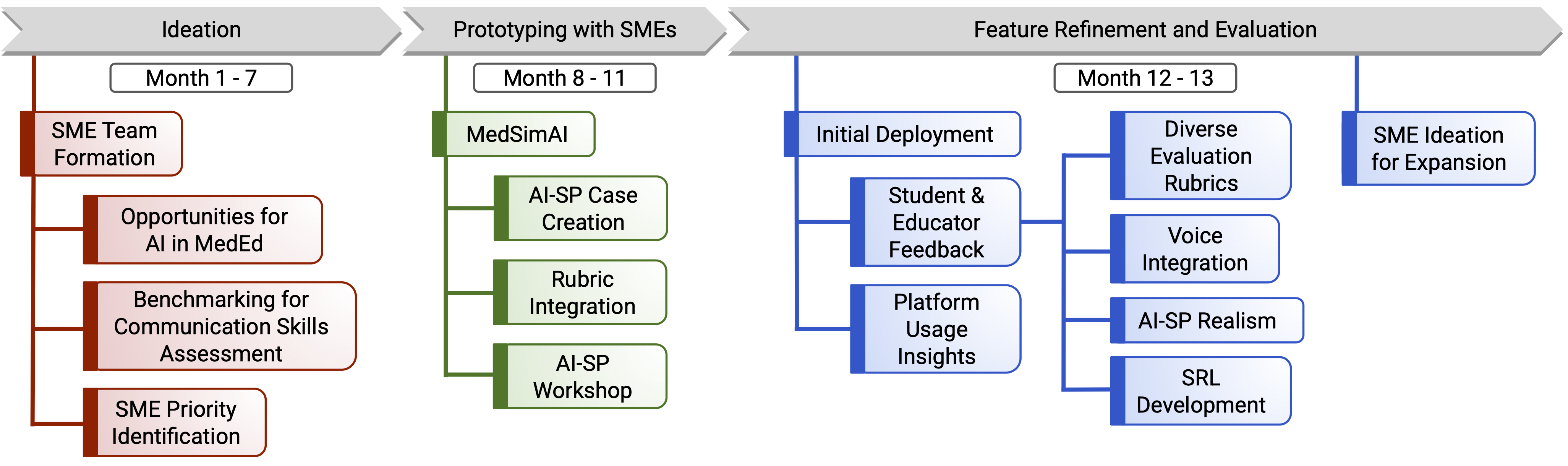}
    \caption{\footnotesize The \emph{MedSimAI} design and development timeline, occurring in three phases with continuous SME collaboration.}
    \Description{A three-phase timeline of design and development with subject matter expert involvement.}
    \label{fig:design-flow}
\end{figure}

\emph{MedSimAI} was iteratively developed through a structured co-design process with subject matter experts (SMEs) from three medical institutions. The team consisted of senior medical education leaders (Assistant Dean for Medical Education, multiple program directors), clinical skills and simulation experts (Directors of Clinical Skills Centers, Director of Simulation Curriculum), and specialists in digital health and AI applications (Director of AI in Healthcare Simulation, Director of Digital Health). Team members held faculty appointments ranging from Assistant to Full Professor across specialties including Internal Medicine, Gastroenterology, Pediatrics, and Neurology.

As illustrated in Figure~\ref{fig:design-flow}, our design process engaged SMEs in focused development cycles, with our team first identifying priorities for creating learning moments in medical education, and benchmarking the current capabilities of LLMs in the evaluation of communication skills. This phase of ideation led to our initial prototype of \emph{MedSimAI}, with each expert contributing to specific platform components including case development, feedback mechanisms, assessment frameworks, and AI-SP behavior and management. Following the initial development of our prototype, we conducted a pilot test of \emph{MedSimAI} at participating institutions in the northeastern United States to gather insights from both educators and students.

\subsection{Ideation: Defining SME Priorities in AI-led Medical Simulation}

\subsubsection{Self-Regulated Learning} 
Our expert team emphasized integrating SRL strategies into clinical skills curricula as a crucial opportunity. As noted by one program director: ``Students need to develop habits of self-assessment and goal-setting that will serve them throughout their careers.'' Drawing from successful SRL applications in medical education \cite{van2018self}, experts identified key strategies: goal setting for specific competencies, reflection on AI-SP interactions, progress tracking across frameworks, and strategic case scheduling. They envisioned these serving dual purposes—maintaining engagement through clear progression while developing essential self-regulated learning habits. However, as our evaluation revealed, implementing these features does not guarantee adoption without appropriate curricular scaffolding.

\subsubsection{Evaluation Frameworks for Communication Skills Assessment} 
Our SMEs noted that different clinical scenarios require distinct evaluation approaches, so \emph{MedSimAI} supports multiple assessment rubrics. They observed that while the Master Interview Rating Scale (MIRS) stands as a widely accepted framework for assessing effective, empathetic clinician–patient communication \cite{MIRSManual2005}, specialized tools are sometimes preferable. For example, SSHADESS provides a strengths-focused psychosocial assessment for adolescent care \cite{Ginsburg2014}, whereas the SPIKES protocol is tailored for delivering bad news in oncology and other sensitive contexts \cite{Baile2000}. This feedback influenced our design of a flexible evaluation system accommodating various frameworks while keeping feedback delivery consistent. 

Similar to the rubric-based evaluation frameworks for communication skills assessments, our SMEs drew attention to the important role played by checklist-based assessments in medical education. They emphasized that, unlike the holistic, scale-based evaluations that rubrics provide, checklists enable a binary test of whether medical students incorporate specific, required elements in patient interactions. The clinical skills directors on our team strongly advocated for direct control by educators over checklist customization, as checklist items must be easily adjusted based on course objectives or encounter types. As both checklists and evaluation frameworks form the foundation of constructive student feedback, our expert team stressed that instructor control over checklists would facilitate alignment between the feedback given by human evaluators (including SPs, coaches, and medical educators) and \emph{MedSimAI}'s automated assessments.

\subsection{Translating Prototyping Insights into Feature Refinement}

Building on SME priorities, we created an initial \emph{MedSimAI} prototype and conducted pilot testing. For one U.S. medical school pilot, we recruited 88 of 179 eligible first-year students. Feedback was gathered through baseline surveys (n=85), post-use surveys (n=19).

Student feedback emphasized practical needs. A \textbf{voice-enabled interface} emerged as top priority: ``Typing feels unnatural—real patients don't communicate through chat'' (MS1 student). Students noted voice interaction would better prepare them for physical OSCEs and reduce cognitive load during practice.

Medical educators, on the other hand, focused on improving the system's educational value through greater customization. Although they reaffirmed the importance of flexible evaluation frameworks and checklists, they emphasized the need for greater \textbf{AI-standardized patient (AI-SP) controllability} to create more realistic training scenarios. Specifically, they highlighted the importance of regulating the AI-SP's volunteered information and language use. This insight led to the development of an instructor dashboard that gives educators precise control over not only evaluation frameworks and checklists but also AI-SP features like behavior, vital signs, and vocabulary.

Medical educators also highlighted a need for a \textbf{professional environment for SRL}, favoring practical clinical approaches over gamified SRL elements. Their feedback shaped several key features. For a strategic planning component, one educator proposed implementing a calendar system that uses clinically relevant terminology, like having students ``add appointments'' rather than ``schedule practice cases.'' For progress tracking, another educator suggested aligning the rubric-based summaries of student engagements with the Kalamazoo essential elements \cite{makoul2001essential}, which categorize clinical encounters into thematic phases. This approach would enable averaging rubric scores by shared essential elements rather than computing global averages. Across all discussions, educators showed uniform interest in maintaining a professional focus in the SRL environment, highlighting the importance of not just what SRL techniques we implemented, but how we fundamentally designed and presented them.

\subsection{Future Directions for Platform Enhancement}
Post-pilot expert discussions identified key expansion opportunities. Longitudinal cases enabling patient follow-up across sessions would support progressive skill development. Integration of additional LCME-required competencies—such as interprofessional teamwork, patient safety and quality-improvement practice, culturally responsive communication, and handoff skills—would strengthen curricular alignment. Finally, experts emphasized gathering detailed student perspectives on SRL feature usage to optimize engagement strategies.

\section{\emph{MedSimAI} Architecture}

Our co-design process yielded \emph{MedSimAI}'s current architecture, comprising three integrated components supporting scalable history-taking and clinical communication skills development.

\subsection{AI Standardized Patient (AI-SP) Simulation}

\emph{MedSimAI}'s AI-SP engine uses instructor-provided case parameters to generate consistent, realistic patient interactions. Instructors customize patient characteristics—demographics, medical history, communication style, emotional state—through structured templates. These parameters feed into carefully crafted system prompts (template examples in \osfRepo) that guide LLM behavior while maintaining clinical authenticity.

The system leverages GPT-4o \cite{openai_gpt4o} for text interactions and OpenAI's Realtime API \cite{openai_realtime_api} for voice conversations. To ensure authentic patient behavior, prompts explicitly instruct the AI to avoid medical jargon, express symptoms in lay terms, and maintain consistent personality traits throughout encounters. Educators can configure cases to be up to 30 minutes long to mirror typical clinical encounters \cite{neprash2023association}, but student feedback suggests that longer durations may be desirable for practice.

\subsection{Assessment and Feedback Framework}

\begin{figure}
    \centering
    \includegraphics[width=\linewidth]{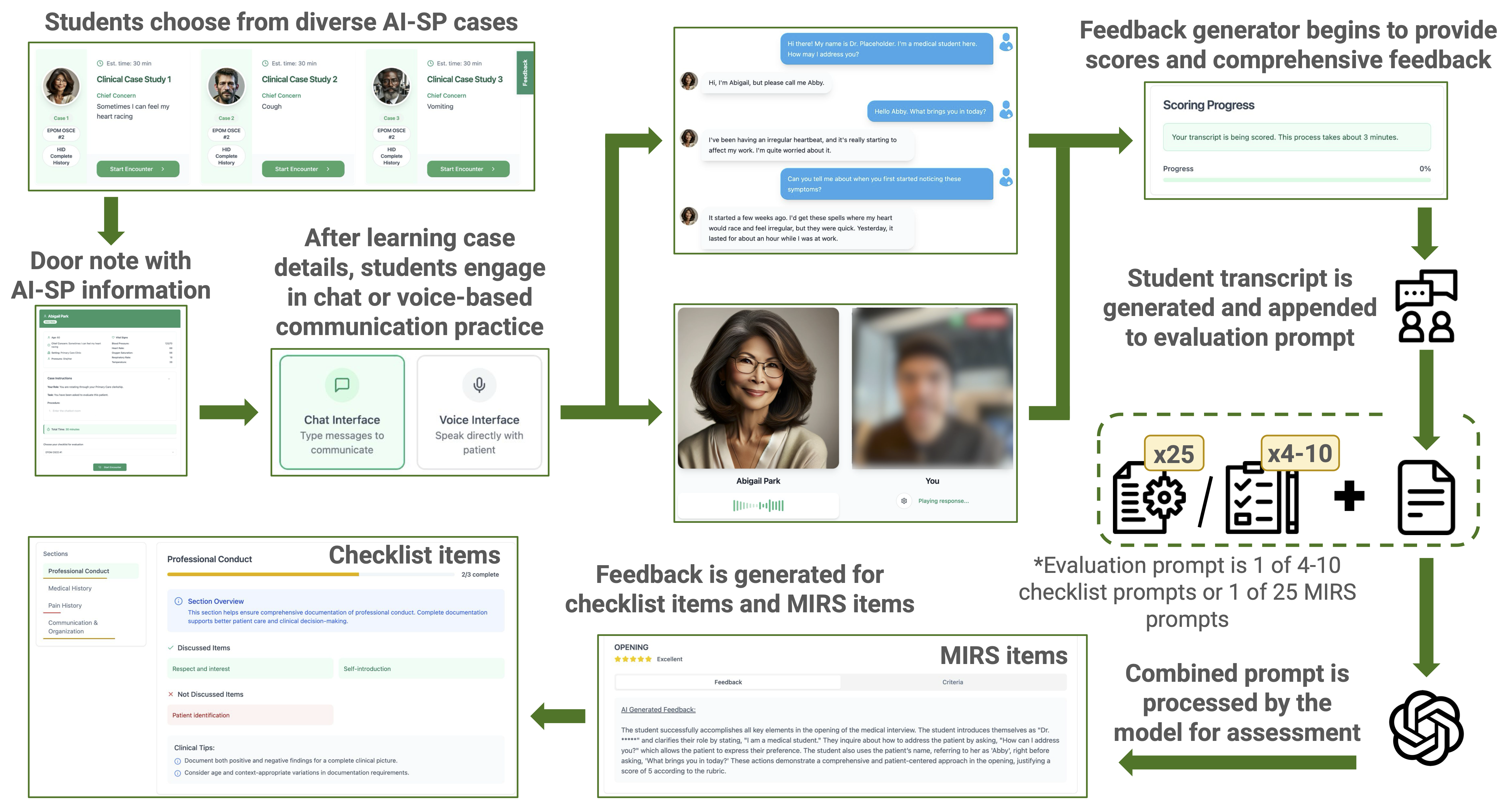}
    \caption{\footnotesize \emph{MedSimAI} workflow: Students select cases, interact via chat or voice, and receive automated feedback based on multiple assessment frameworks. Not pictured are the SRL-enhancing components shown in Figure~\ref{fig:learning-hub}.}
    \Description{Workflow diagram from case selection through interaction and automated feedback.}
\label{fig:ai-sp}
\end{figure}

\emph{MedSimAI} processes completed encounters using specialized evaluation prompts (examples in \osfRepo). Each case can attach one or more assessment modules: rubric-based ratings (e.g., the 28-item Master Interview Rating Scale (MIRS) on a 1--5 scale) and case-specific checklists that encode required history-taking content (e.g., red-flag symptoms to assess).

For rubrics, items are scored independently using item-specific prompts that include anchored criteria and require quote-grounded justifications; outputs are returned as structured JSON for reliable parsing and aggregation. For checklists, each item is evaluated as assessed/not assessed (and present/absent/unknown when applicable) with supporting excerpts; checklist completion is computed as the proportion of required items assessed. Educators can select existing frameworks (e.g., SPIKES, SSHADESS) or author custom checklists via the instructor dashboard.

Learners receive an itemized report highlighting strengths and gaps with evidence quotes and suggestions, alongside aggregated views mapped to clinically familiar categories (Kalamazoo essential elements) for longitudinal tracking in the Learning Hub. Assessment typically completes within 2--3 minutes post-encounter, enabling rapid formative feedback.

\subsection{Learning Hub: A Self-Regulated Learning Interface}

\begin{figure}
    \centering
    \includegraphics[width=0.9\linewidth]{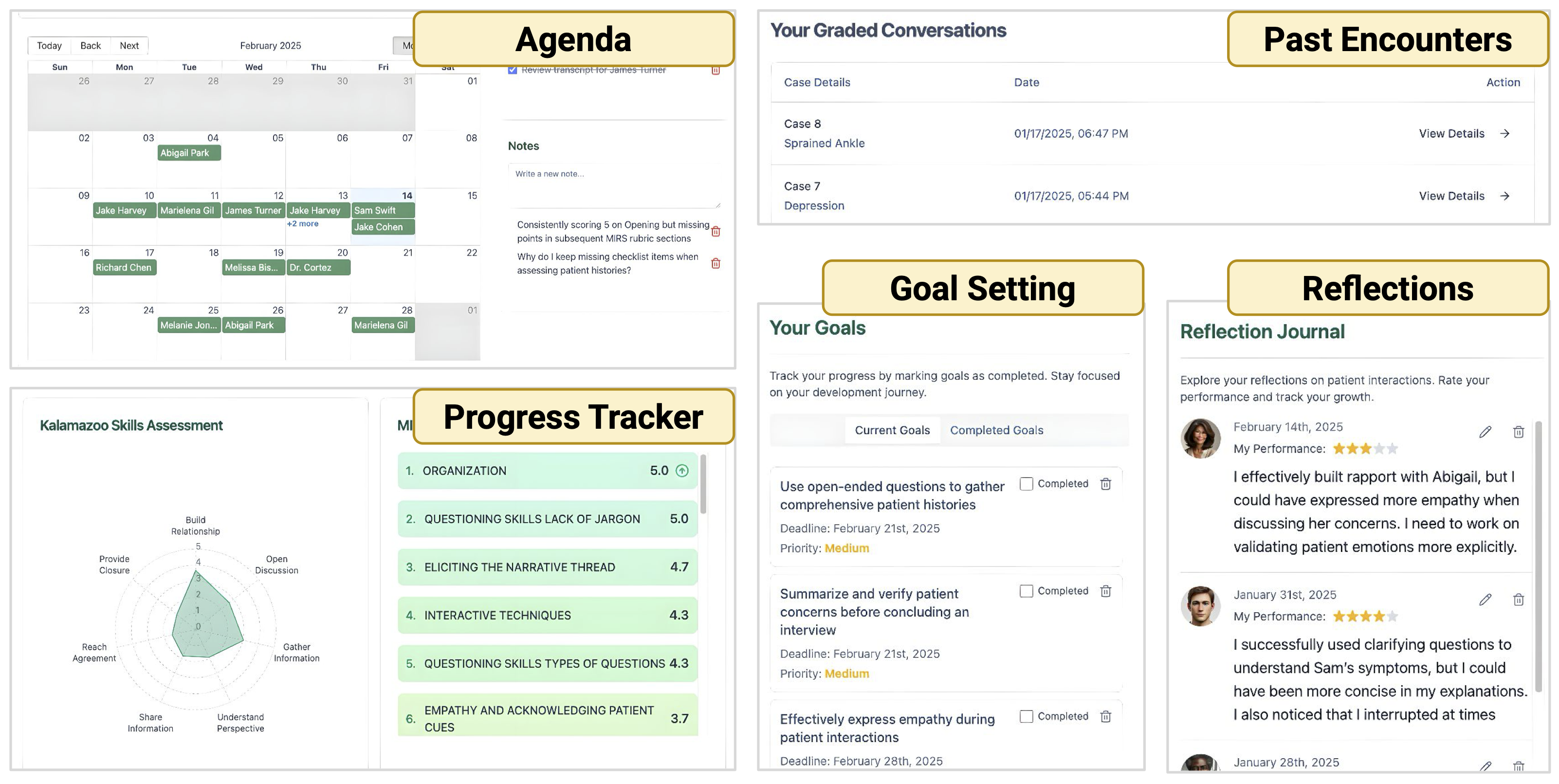}
    \caption{\footnotesize Learning Hub integrates five SRL components using clinical metaphors.}
    \Description{Dashboard-like interface showing scheduling, charts, progress, goals, and reflections.}
    \label{fig:learning-hub}
\end{figure}

The Learning Hub (Figure \ref{fig:learning-hub}) implements evidence-based SRL strategies through clinically-authentic interfaces. Components include: appointment scheduling (strategic planning), patient chart reviews (past encounters), competency dashboards (progress tracking), clinical goal setting, and encounter reflections. The Progress Tracker aggregates performance using Kalamazoo elements, providing clinically meaningful skill categories rather than abstract scores. Despite careful design following expert input, usage data revealed limited engagement with these features, suggesting need for stronger curricular integration.

\section{Evaluation}

\subsection{Method}
We conducted a multi‑site deployment and evaluation to examine:  
(1) platform‑usage patterns across institutions,  
(2) clinical‑performance impact where measurable,  
(3) student perceptions and perceived value, and  
(4) the reliability of \emph{MedSimAI}’s automated scoring relative to human evaluators.

\subsubsection{Participants and Settings} 
The platform was deployed at three U.S. medical schools: a major private research-intensive medical school in a large urban center (Institution A), a large West Coast public institution (Institution B), and an Ivy League-affiliated private institution in the Northeast (Institution C). All sites obtained IRB approval to analyse de-identified data, and each made an optional survey available to students who tried the tool. Institution B embedded that survey in a structured pilot for its entire first-year class. During the pilot’s first session, students who completed at least one encounter could fill out a baseline questionnaire administered immediately afterward. A follow-up questionnaire delivered several weeks later yielded additional responses. Institutions A and C also circulated the survey but collected fewer than five completed forms apiece, an insufficient number for quantitative analysis.

\subsubsection{Data Collection}
We drew on four complementary data sources. 
\textbf{(1) Platform telemetry:} Across the three sites, we analyzed \emph{1{,}024} completed \emph{MedSimAI} encounters from \emph{410} unique learners. For each session we logged modality (voice vs.\ text), conversation duration (minutes), number of dialogue turns, case ID/version, and automated assessment outputs (checklist completion \% and MIRS item- and overall scores). Session‑level self‑regulated learning (SRL) artifacts (student goals and free‑text reflections) were also captured; in total we analyzed 840 reflections for the thematic analysis reported in Table~\ref{tab:reflection-themes}, and we derived reflection length (characters) as a coarse SRL proxy for telemetry models. All quantitative analyses used sessions with available automated scoring (sessions aborted before feedback generation were excluded).
\textbf{(2) Surveys:} At Institution~B, students completed a baseline questionnaire (confidence and SRL behaviors; $n{=}85$) and an exit questionnaire ($n{=}19$) consisting of Likert-type items (7-point for exam anxiety/preparation/value; 5-point for feedback quality/usability) plus four open-ended prompts.
\textbf{(3) Course performance:} We obtained OSCE data from two sites. Institution~A provided history-taking scores for a prior cohort without \emph{MedSimAI} access ($n{=}100$) and the first cohort with access ($n{=}100$). Institution~B provided Demonstration of Clinical Skills (DOCS) results for voluntary \emph{MedSimAI} participants ($n{=}88$) and non-participants ($n{=}91$). 
\textbf{(4) Validation corpus:} To benchmark automated scoring, we rescored \emph{104} first-year OSCE transcripts from Institution~C that had been previously rated on the 28-item Master Interview Rating Scale (MIRS) by trained external evaluators; seven items not applicable to the two cases and two items lacking audio were excluded, yielding \emph{19} items per transcript for analysis.
\textbf{Qualitative data and organization:} Open-ended survey responses formed two cohort-specific corpora. For first-year learners (Complete Hx OSCE), we analyzed 38 overall impressions, 28 most-valuable features, 24 areas for improvement, and 13 desired functionalities. For third-year learners (10-case OSCE), we analyzed 14 overall impressions, 14 most-valuable features, 13 areas for improvement, and 7 desired functionalities. We conducted an inductive thematic analysis and consolidated themes in a cross-cohort summary to support interpretation in the Results.

\subsubsection{Analysis}
Survey responses were analyzed using descriptive statistics (e.g., means and standard deviations for Likert‑scale items) and qualitative synthesis of open‑ended feedback. For the validation study, \emph{MedSimAI} re‑scored each transcript on the MIRS; seven items not relevant to the two cases and two items lacking audio were omitted, leaving 19 items per transcript. Following prior benchmarking methodology \cite{geathers2025benchmarking}, we compared AI and human scores using three accuracy metrics: (1) Exact Accuracy (strict), (2) Off‑by‑One Accuracy (moderate), and (3) Thresholded Accuracy (lenient; 1–2 vs.\ 3–5). 

We also ran exploratory models on platform telemetry (1{,}024 sessions; 410 learners; 30 cases) to examine institutional and case effects and usage–performance relationships. We used Kruskal–Wallis tests with Dunn post‑hoc comparisons for group-wise differences and linear mixed‑effects models (random intercepts for learner with variance components for case ID) to estimate adjusted associations for institution, case ID, session count, modality (voice vs.\ text), dialogue turns, and reflection length vs.\ next‑session change in MIRS. Full model specifications and outputs are available in our anonymized OSF repository (\osfRepo).

\subsection{Findings}

\subsubsection{Usage Patterns and Engagement}
Table \ref{table:usage-by-site} shows engagement by site. Averaged across all users, students completed 2.50$\pm$2.89 cases, but this masked wide variation: Institution B recorded the highest practice intensity (3.48$\pm$3.58 cases; 73.1\% completing $\geq 2$ cases), whereas Institution A had broader but lighter uptake (2.25$\pm$1.75 cases; 57.8\% completing $\geq 2$ cases). Institution C showed minimal engagement (1.35$\pm$0.48 cases; 29.0\% completing $\geq 2$ cases). Overall, 59.5\% of learners completed multiple encounters.

A small subpopulation of students exhibited exceptional engagement: 12 students (2.9\%) completed 8 or more cases, accounting for 124 total encounters (12.1\% of all platform usage). The top performer completed 20 cases, representing 2\% of all platform activity alone. High-engagement patterns varied substantially by institution: Institution B had 9 high-engagement students (8.7\% of their cohort), Institution A had 3 (1.1\%), and Institution C had none (0\%).

\begin{table*}[t]
\centering
\caption{Platform engagement and performance statistics by institution and overall}
\label{table:usage-by-site}
{%
\begin{tabular}{lcccc}
\toprule
 & \textbf{Inst.\,A} & \textbf{Inst.\,B} & \textbf{Inst.\,C} & \textbf{Overall} \\
\midrule
Students (\textit{n}) & 275 & 104 & 31 & 410 \\
Total cases completed           & 620 & 362 & 42 & 1,024 \\
\midrule
\multicolumn{5}{l}{\textbf{Engagement Metrics}} \\
Cases / student & 2.25$\pm$1.75 & 3.48$\pm$3.58 & 1.35$\pm$0.48 & 2.50$\pm$2.89 \\
Students completed: 1 case                            & 116\,(42.2\%) & 28\,(26.9\%) & 22\,(71.0\%) & 166\,(40.5\%) \\
\quad 2+ cases                                    & 159\,(57.8\%) & 76\,(73.1\%) & 9\,(29.0\%) & 244\,(59.5\%) \\
\quad 5+ cases                                   & 15\,(5.5\%) & 26\,(25.0\%) & 0\,(0.0\%) & 41\,(10.0\%) \\
Conversation duration (min.)         & 19.8$\pm$9.1 & 16.9$\pm$7.0 & 12.7$\pm$7.5 & 18.5$\pm$8.5 \\
Dialogue turns per case       & 38.0$\pm$19.9 & 39.4$\pm$14.6 & 25.7$\pm$16.4 & 38.0$\pm$18.2 \\
Voice‑only users                                     & 98\,(35.6\%) & 56\,(53.8\%) & 18\,(58.1\%) & 172\,(42.0\%) \\
Text‑only users & 137\,(49.8\%) & 30\,(28.8\%) & 12\,(38.7\%) & 179\,(43.7\%) \\
Both modalities                                      & 40\,(14.5\%) & 18\,(17.3\%) & 1\,(3.2\%)  & 59\,(14.4\%) \\
\midrule
\multicolumn{5}{l}{\textbf{Performance Metrics}} \\
MIRS overall score & 3.57$\pm$0.70 & 3.69$\pm$0.60 & 3.18$\pm$0.54 & 3.60$\pm$0.66 \\
Checklist completion (\%) & 54.8$\pm$16.0 & 59.3$\pm$21.5 & 26.5$\pm$16.4 & 55.2$\pm$19.2 \\
\bottomrule
\end{tabular}}
\end{table*}

\subsubsection{Clinical Performance Results}

\begin{figure}[t]
  \centering
  \includegraphics[width=0.6\linewidth]{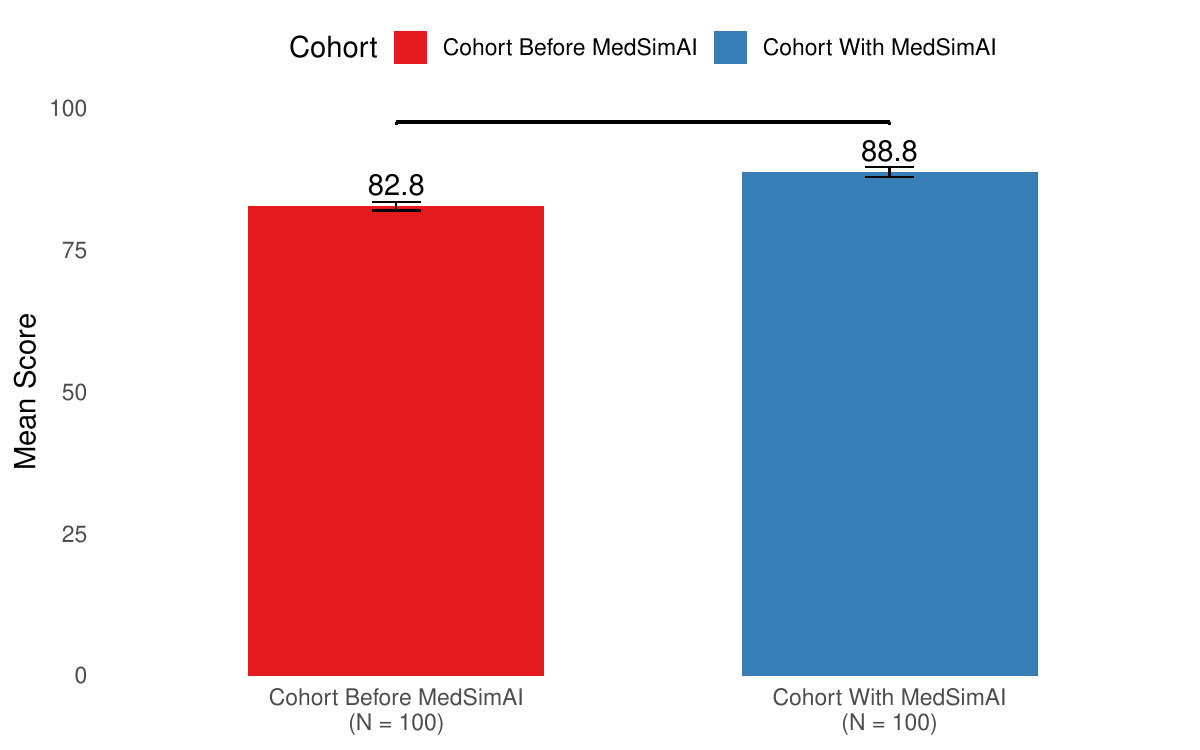}
  \vspace{4pt}
  \caption{OSCE history-taking performance at Institution~A. Bars show mean scores with standard deviations (Pre-platform $n=100$: $82.8\pm7.6$; Post-platform $n=100$: $88.8\pm8.5$). Difference $=6.0^{***}$, effect size $d=0.75$. $^{***}\!p<0.001$ (two-sided Welch $t$-test).}
  \label{fig:osce-performance}
\end{figure}

\paragraph{Institution\,A (quasi‑experimental cohort comparison)}
An uncontrolled comparison of OSCE history‑taking scores showed a statistically significant improvement after the adoption of \emph{MedSimAI} (Figure~\ref{fig:osce-performance}).  
The cohort with platform access scored 6.0 points higher ($p<0.001$), corresponding to a large effect size ($d=0.75$).  
Although multiple confounders may have influenced this outcome, local instructors noted that the platform’s introduction was the most substantial curricular change between the two cohorts.

\begin{figure}[t]
\centering
\caption{DOCS performance, Institution\,B (88 volunteers vs.\ 91 non‑participants)}
\includegraphics[width=0.8\linewidth]{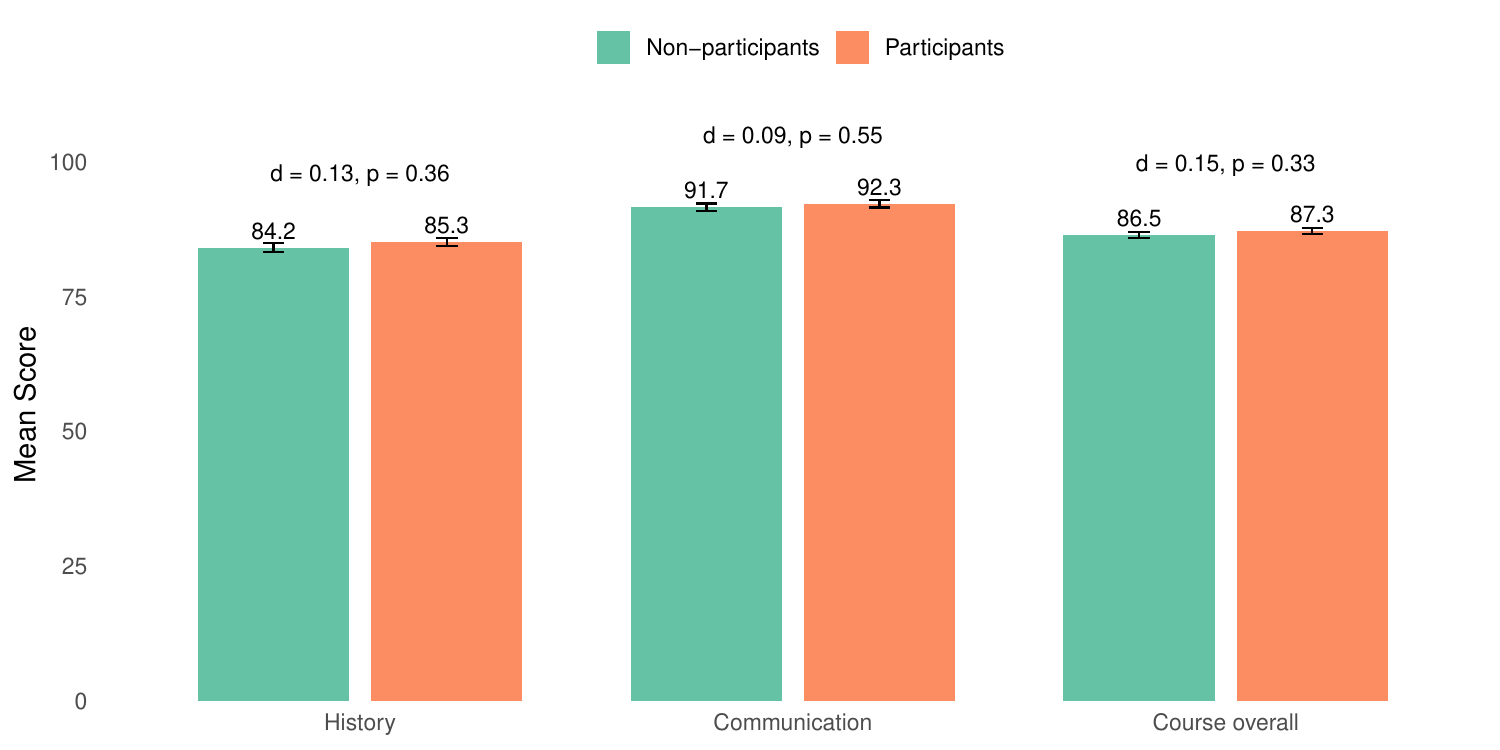}
\label{fig:docs}
\end{figure}

\paragraph{Institution\,B (self‑selected pilot participants)}
Eighty‑eight of 179 students volunteered to use \emph{MedSimAI} in the weeks preceding their Demonstration of Clinical Skills (DOCS) examination; the remaining 91 students served as a non‑participant comparison group.  
Because \emph{MedSimAI} practice cases focused on history‑taking, communication, and diagnostic reasoning, we did not expect changes in DOCS domains unrelated to those skills (e.g., physical exam).  
Welch $t$‑tests revealed no statistically significant differences on the target domains (all $p>0.30$; Figure \ref{fig:docs}), suggesting that self‑directed use in this context did not translate into measurable exam gains.

\subsubsection{Independent Validation of Automated Scoring}\label{sec:valid}
The \emph{MedSimAI} scoring engine achieved 32.5\% exact accuracy, 64.1\% off-by-one accuracy, and 87.0\% thresholded accuracy. While exact score matches occurred in only one-third of cases, \emph{MedSimAI} correctly separated under-performing from proficient learners in 87\% of encounters and agreed with human raters within a single point two-thirds of the time. In practice, that level of concordance is more than adequate for formative triage—flagging students who need remediation and relieving faculty of the need for first-pass scoring. Exact alignment will still require human oversight, but these results show the system can already handle the high-volume, low-stakes portion of the feedback workflow.

\subsubsection{Student experience and perceived value}
Post‑use survey data from Institution~B (\textit{n}=19) painted a generally positive—yet nuanced—picture. On 7‑point Likert items (1=\emph{strongly disagree}, 7=\emph{strongly agree}), mean ratings were: “reduced exam anxiety” $M=5.38$, “improved preparation” $M=5.38$, and “supported my learning goals” $M=5.15$. Feedback quality items on 5‑point scales averaged 4.38 (\emph{constructive}), 4.46 (\emph{actionable}), and 4.31 (\emph{helped me improve}). 

Thematic analyses of open‑response items across cohorts identified three consistent themes: value of detailed, quote‑based feedback; technical glitches; and perceived lack of realism. Representative comments included “\emph{The feedback is extremely detailed, and features direct quotes}” and “\emph{It was helpful, but a bit glitchy}.” %
Third‑year students echoed these themes and added concerns about scoring transparency and case complexity (e.g., “\emph{It was glitchy — I did a whole 30 min sim and couldn’t access my graded report}”; “\emph{The cases were MUCH simpler than the OSCE cases}”). %
A cross‑cohort summary highlights a developmental pattern: first‑years emphasized confidence/flow and basic realism, whereas third‑years emphasized differential diagnosis, physical‑exam integration, and rubric‑aligned scoring. %

\subsubsection{Thematic analysis of learner reflections}
To understand how learners self‑assess after practice, we conducted an inductive thematic analysis \cite{braun2006using} of 840 free-text reflections written immediately after sessions in the platform’s SRL interface. Nine salient themes emerged (Table~\ref{tab:reflection-themes}). The most frequent were \emph{missed/forgotten items} (26.3\%) and \emph{organization/flow} (23.0\%), followed by \emph{review of systems (ROS) \& history‑taking technique} (22.4\%) and \emph{empathy/rapport \& caregiver management} (18.3\%); learners also surfaced \emph{differential/clinical reasoning} (13.2\%) and \emph{time/logistics \& tech issues} (12.4\%). These reflections align with survey comments that emphasize timing pressure, realistic voice interaction, and the value of structured feedback, while highlighting tensions between checklist completeness and conversational flow (\S\ref{sec:valid}). Representative quotes included: “\emph{Forgot to ask ROS questions},” “\emph{Transitions could be improved},” “\emph{Didn’t know how to console the mother in such a stressful situation},” and “\emph{Diagnosing was hard—wish I knew more medicine}.” 

\begin{table*}[t]
\centering
\caption{Thematic coding of 840 post-simulation session reflections.}
\label{tab:reflection-themes}
\small
\setlength{\tabcolsep}{4pt}
\begin{tabular}{@{}p{0.30\linewidth}rrp{0.55\linewidth}@{}}
\toprule
\textbf{Theme} & \textbf{Count} & \textbf{\%} & \textbf{Example reflection} \\
\midrule
Missed/forgotten items & 221 & 26.3\% & “I forgot to ask if she uses any supplements.” \\
Organization \& flow of interview & 193 & 23.0\% & “I wish I had asked questions in a more systematic way.” \\
ROS \& history-taking technique & 188 & 22.4\% & “Needed more time for the ROS; I rushed it.” \\
Empathy/rapport \& caregiver management & 154 & 18.3\% & “I didn’t know how to console the mother in such a stressful situation.” \\
Differential/clinical reasoning & 111 & 13.2\% & “Once a fever was mentioned, I got anchored too early.” \\
Time \& logistics/tech issues & 104 & 12.4\% & “The mic kept disconnecting during the session.” \\
Third-party/limited communication & 96 & 11.4\% & “It was weird only talking to the parent instead of the patient.” \\
Knowledge/confidence gaps & 47 & 5.6\% & “I’m not well versed in taking a focused neuro history.” \\
Checklist vs.\ conversation tension & 33 & 3.9\% & “The interview felt too bullet-point oriented.” \\
\bottomrule
\end{tabular}
\end{table*}

\paragraph{Exploratory cross‑site and case‑level performance (platform‑scored).}
Across 1{,}024 platform‑scored encounters we observed strong institutional differences in both checklist completion and MIRS (Kruskal–Wallis $p<0.001$; all Dunn pairwise tests significant). Mixed‑effects models (random intercept for case) showed higher checklist completion at Institution~B vs.\ Institution~A (+4.05 percentage points, $p=0.001$) and substantially lower completion at Institution~C (–28.69, $p<0.001$); for MIRS, Institution~A was higher (+0.147, $p=0.001$) and Institution~C lower (–0.354, $p=0.001$). Case ID also predicted both outcomes (Kruskal–Wallis $p<0.001$), indicating non‑trivial variation in case difficulty. 

We next examined usage–performance relationships. After adjusting for institution and case, the number of sessions completed was \emph{not} associated with checklist completion ($\beta\approx-0.003$, $p=0.992$) but showed a small positive association with MIRS overall ($\beta\approx0.019$ points per session, $p=0.032$). Modality did not predict checklist completion but voice sessions had slightly lower MIRS than text (mixed‑effects $\beta\approx-0.132$, $p=0.001$). The number of dialogue turns per encounter was positively associated with checklist completion ($\beta\approx0.566$ percentage‑points per turn, $p<0.001$) but not with MIRS ($p=0.687$). Finally, reflection length did not predict change in next‑session MIRS ($p=0.258$). Collectively, these exploratory models suggest that context (site, case) and within‑session conduct (turn‑taking) are more predictive of rubric coverage than raw practice dose, while communication quality (MIRS) shows a small but detectable relation to repeated practice.

\section{Discussion and Limitations}

Our multi‑institutional deployment demonstrates \emph{MedSimAI}’s potential for enhancing history‑taking and clinical‑communication training while revealing important implementation challenges.  
The significant improvement at Institution A ($d=0.75$) suggests that intensive, curriculum‑integrated use can translate into performance gains, whereas the absence of a similar effect at Institution B—where participation was voluntary—underscores the importance of contextual factors (\emph{we discuss potential reasons for this discrepancy below}).  Faculty identified \emph{MedSimAI} as the primary curricular change between cohorts at Institution A, but other factors cannot be ruled out.

\paragraph{Summary of key findings.}
Across three medical schools we observed:  
(1) a large, quasi‑experimental performance gain at Institution A;  
(2) null DOCS differences at Institution B despite positive learner attitudes;  
(3) automated scoring that identified proficiency thresholds with 87\% accuracy; and  
(4) sustained, voluntary use by 59.5\% of students, totalling 1,024 simulated encounters.

\subsubsection{Implementation and curricular integration} 
Success requires more than technical deployment. Fixed 30-minute sessions poorly served OSCE preparation at Institution B, where exams allocate 20 minutes for both history and physical examination. Voice interface challenges limited adoption despite student preference for verbal interaction. Most critically, SRL features saw minimal engagement without curricular integration—suggesting that providing tools alone is insufficient for behavior change.

The contrasting performance results between institutions highlight the importance of implementation context. Institution A's significant improvement may reflect stronger curricular integration or a different student population compared to Institution B's voluntary pilot, which shows minimal differences.

\subsubsection{Closing the analytics loop: findings to action}
Beyond reporting outcomes, we used telemetry and qualitative data to inform concrete design and implementation changes, making the learning-analytics cycle explicit. Table~\ref{tab:analytics-loop} summarizes representative ``finding $\rightarrow$ action'' pairs that emerged from deployment across sites, cases, and modalities.

\begin{table*}[t]
\centering
\caption{Analytics findings and resulting design/curriculum actions (implemented or prioritized for subsequent iterations; illustrative examples).}
\label{tab:analytics-loop}
{\small
\setlength{\tabcolsep}{4pt}
\renewcommand{\arraystretch}{1.15}
\begin{tabular}{@{}c p{0.43\linewidth} p{0.43\linewidth}@{}}
\toprule
\textbf{\#} & \textbf{Finding (data)} & \textbf{Action (design/curriculum)} \\
\midrule
1 & Curriculum-embedded deployment at Institution~A coincided with a +6.0 point OSCE gain, whereas voluntary participation at Institution~B showed no measurable DOCS differences. & Informed a shift toward curriculum-embedded use with structured practice schedules rather than optional access. \\
\addlinespace[0.6ex]
2 & Telemetry showed limited engagement with SRL features without explicit curricular embedding. & Motivated redesign priorities to integrate SRL steps (goal setting, reflection) into coursework and encounter flow. \\
\addlinespace[0.6ex]
3 & Fixed 30-minute sessions mismatched local OSCE timing and learner feedback. & Motivated adjustable encounter durations and timing presets aligned to local assessment formats. \\
\addlinespace[0.6ex]
\addlinespace[0.6ex]
4 & Strong case-level effects were observed in checklist completion and MIRS. & Motivated recalibration of case difficulty and sequencing into progressive practice tiers. \\
\addlinespace[0.6ex]
5 & Automated scoring aligned well on proficiency thresholds (87\% thresholded) but less on exact scores (32.5\%). & Positioned automated scoring as formative triage with uncertainty framing and guidance for human review. \\
\bottomrule
\end{tabular}}
\end{table*}

\subsubsection{Limitations} 
Several limitations qualify our findings. The quasi-experimental design comparing historical cohorts at Institution~A cannot definitively establish causation due to potential confounders (e.g., curriculum changes, instructor variations, cohort differences). While faculty identified \emph{MedSimAI} as the primary curricular change, other factors cannot be ruled out. Survey data come primarily from one institution (Institution~B) with modest response rates. Performance comparisons use different metrics across institutions, and we do not isolate the causal contribution of individual platform components (simulation, scoring, or SRL scaffolds) because the study evaluates the integrated system. Finally, we did not instrument fine-grained feedback uptake behaviors (e.g., time spent viewing feedback or revisions following feedback), limiting mechanism-oriented modeling beyond coarse proxies (e.g., reflection length). LLM inconsistencies (e.g., overly formal language) and ASR instability also affected perceived realism.

\subsubsection{Future Directions} 
Addressing identified limitations requires both technical and pedagogical innovation. Technical priorities include adjustable encounter durations, improved voice recognition, multilingual AI-SPs, and enhanced prompt engineering for realism. Pedagogical priorities include mandatory SRL exercises within coursework, structured skill progression curricula, and longitudinal outcome measurement. The independent validation results suggest focusing on formative rather than summative assessment applications.

\section{Conclusion}
\emph{MedSimAI} demonstrates that carefully designed AI-powered simulation shows promise for enhancing medical students' clinical skills, as suggested by performance gains at one institution, though confounders limit definitive causal claims. Through systematic co-design with medical educators and multi-institutional deployment, we show that LLM-based standardized patients offer scalable solutions to longstanding medical education challenges. However, our findings also reveal that technical innovation alone is insufficient—successful implementation requires thoughtful curricular integration, addressing practical constraints like timing flexibility, and explicit support for self-regulated learning behaviors. As medical education continues evolving to meet 21st-century healthcare needs, AI-powered tools like \emph{MedSimAI} can play a crucial role in ensuring all students receive high-quality clinical skills training, regardless of institutional resources. Exploratory mixed‑effects models highlight strong institution and case effects and only a small association between repeated use and MIRS improvements, indicating that how and what students practice matters more than raw volume. Designing for staged complexity and curricular integration may therefore yield larger downstream gains than unguided, optional use.

\begin{acks}
\textbf{Declaration of interest.} The authors declare no competing interests.

\textbf{Funding.} This research did not receive any specific grant from funding agencies in the public, commercial, or not-for-profit sectors.

\textbf{Use of generative AI in writing.} During the preparation of this work, the authors used ChatGPT (OpenAI) solely for grammar and readability improvements. After using this tool, the authors reviewed and edited the content and take full responsibility for its integrity.
\end{acks}

\bibliographystyle{ACM-Reference-Format}
\bibliography{bibliography}

\end{document}